\documentclass[twocolumn,journal]{IEEEtran}

\usepackage{graphicx}
\usepackage{amsmath}
\usepackage{urwchancal}
\usepackage[mathscr]{eucal}

\usepackage{amssymb}
\usepackage{comment}
\usepackage{tabu}
\usepackage{epsfig}
\usepackage{algorithmic}
\usepackage{algorithm}

\usepackage{amsthm}
\usepackage{cite}
\usepackage{float}
\usepackage{color}
\usepackage{balance}

            \DeclareSymbolFontAlphabet{\mathrm}    {operators}
\DeclareSymbolFontAlphabet{\mathnormal}{letters}
\DeclareSymbolFontAlphabet{\mathcal}   {symbols}
\DeclareMathAlphabet      {\mathbf}{OT1}{cmr}{bx}{n}
\DeclareMathAlphabet      {\mathsf}{OT1}{cmss}{m}{n}
\DeclareMathAlphabet      {\mathit}{OT1}{cmr}{m}{it}
\DeclareMathAlphabet      {\mathtt}{OT1}{cmtt}{m}{n}
\DeclareMathAlphabet{\mathpzc}{OT1}{pzc}{m}{it}
\begin{document}
\title{\vspace{-0.5cm}\LARGE An Artificial-Noise-Aided Hybrid TS/PS \\ Scheme for OFDM-Based SWIPT Systems}
\author{Ahmed El Shafie, Kamel Tourki, Naofal Al-Dhahir \vspace{-1.6\baselineskip}
\begin{tabular}{c}
\small $^\dagger$Electrical Engineering Dept., University of Texas at Dallas, USA. \\
\small $^\star$Mathematical and Algorithmic Sciences Lab,
France Research Center, Huawei Technologies Co. Ltd.
\end{tabular}
\thanks{This paper was made possible by NPRP grant number  8-627-2-260 from the Qatar National Research Fund (a member of Qatar Foundation). The statements made herein are solely the responsibility of the authors.}}
\date{}
\maketitle
\begin{abstract}
We propose a new artificial-noise aided hybrid time-switching/power-splitting scheme for orthogonal frequency-division multiplexing (OFDM) systems to securely transmit data and transfer energy to a legitimate receiving node. In our proposed scheme, the cyclic prefix has two more benefits in addition to the cancellation of the inter-symbol interference between the OFDM blocks. Firstly, it enables the legitimate transmitter to send artificial-noise (AN) vectors in a way such that the interference can be canceled at the legitimate receiver prior to information decoding. Secondly, its power is used to energize the legitimate receiver. We optimize the cyclic prefix length, the time-switching and power-splitting parameters, and the power allocation ratio between the data and AN signals at the legitimate transmitter to maximize the average secrecy rate subject to a constraint on the average energy transfer rate at the legitimate receiver. Our numerical results demonstrate that our proposed scheme can achieve up to $23\%$ average secrecy rate gain relative to a pure power-splitting scheme.
\end{abstract}
\begin{IEEEkeywords}
Energy harvesting, OFDM, security, SWIPT.
\end{IEEEkeywords}
\vspace{-0.4cm}
\section{Introduction}
\vspace{-0.1cm}
Radio-frequency (RF) energy harvesting (EH) is a powerful technology that enables wireless nodes to be charged using the ambient RF transmissions. A wireless node converts the received RF
transmissions into direct current (DC) electricity. Most existing research work adopts either time-switching (TS) simultaneous wireless
information and power transfer (SWIPT) schemes \cite{zhang2013mimo}, where the receiver switches between data reception and energy harvesting, or power-splitting (PS) SWIPT schemes \cite{zhang2013mimo}, where the receiver splits the signal into two streams of different powers
for decoding information and harvesting energy separately to
enable simultaneous information decoding and energy harvesting.

Unlike the commonly-used pure TS or pure PS schemes for SWIPT, we propose a hybrid TS/PS scheme for wireless orthogonal frequency-division multiplexing (OFDM) systems. In \cite{ng2013wireless}, the authors investigated orthogonal frequency-division
multiple-access (OFDMA) systems with SWIPT. Resource allocation
schemes were proposed for the maximization of the energy efficiency of data
transmissions. The implementation of the schemes in \cite{ng2013wireless} is challenging since the authors assumed that the PS ratio can be different across the OFDM sub-channels. As reported in \cite{zhou2014wireless}, in practical circuits, (analog) power splitting is performed before (digital) OFDM demodulation. Thus, for an OFDM-based SWIPT system, all sub-channels need to be power split with the same power ratio.
In \cite{zhou2014wireless}, the authors considered time-division multiple-access (TDMA) and OFDMA schemes. Each receiver applies
either TS or PS to coordinate
the EH and information decoding processes. In \cite{poor}, the authors proposed new schemes to extend
the battery life of an OFDM receiver by exploiting the power in the cyclic prefix (CP) to power an EH receiver.

In \cite{a5}, the authors proposed temporal artificial-noise (AN) injection for the single-input single-output single-antenna eavesdropper (SISOSE) OFDM system in which a time domain AN signal is added to the data signal before transmission. The temporal AN signal is designed to be canceled at the legitimate receiver prior to data decoding. \color{black} In \cite{7378525}, the authors investigated SWIPT in SISO-OFDMA systems under coexistence
of information receivers (IRs) and energy receivers (ERs). The
IRs are served with best-effort secrecy data and the ERs harvest
energy with minimum required harvested power. For each IR, all other receivers
(IRs and ERs) are potential eavesdroppers (Eves). Unlike  \cite{7378525}, we do not assume knowledge of Eves' instantaneous channel state information (CSI) at the legitimate transmitter. The proposed scheme in \cite{7378525} can completely fail in such scenarios since the authors assume that there is a secure way to share secret AN with the legitimate receivers a priori (OFFLINE) and privately/securely from the other nodes in the network, which is not practical in case of no Eve's CSI at the legitimate transmitting node. In \cite{zhang2016energy}, each user applies the PS scheme to coordinate the
energy harvesting and information decoding processes. Without using AN-injection schemes, the authors considered a co-located
SWIPT system by using a PS scheme against eavesdropping.

In this letter, we consider an OFDM scenario with two objectives: 1) securing the legitimate transmissions from passive eavesdropping, 2) transferring energy to the legitimate receiver. The main contributions of this letter are summarized as follows. We propose an efficient AN-aided joint TS/PS SWIPT scheme to power legitimate receiver (Bob) and secure the legitimate transmissions from passive eavesdropping for OFDM-based systems. We exploit the temporal degrees of freedom provided by the CP structure of OFDM blocks to degrade the received signal at Eve and to power Bob. We derive a closed-form expression for the instantaneous secrecy rate of the proposed scheme and show the impact of the PS and TS parameters in addition to other system's parameters (e.g. transmit power, CP length, OFDM block size, etc) on the links' instantaneous rates. Moreover, we derive a closed-form expression for the energy harvested at Bob under our proposed scheme and show the impact of the system's parameters on the energy harvesting rate. We optimize the CP length, the TS and PS parameters, and the power allocation ratio between the data and AN signals at the legitimate transmitter to maximize the average secrecy rate subject to a constraint on average energy transfer rate at~Bob.

\emph{\underline{Notation:}}  Unless otherwise stated, lower- and upper-case
bold letters denote vectors and matrices, respectively. \ $\mathbf{I}_N$ and $\mathbf{F}$ denote, respectively, the identity matrix whose size is $N\times N$ and
the fast Fourier transform (FFT) matrix.\color{black}\ $\mathbb{C}^{M \times N}$ denotes the set of all complex matrices of size $M\times N$. \ $(\cdot)^\top$ and $(\cdot)^*$ denote transpose and Hermitian (i.e. complex-conjugate transpose) operations, respectively.  \color{black}\ $|\cdot|$ denotes the cardinality of a set. $\mathbb{R}^{M \times N}$ denotes the set of real matrices of size $M \times N$. ${[\cdot]_k}$ denotes the $k$th entry of a vector. $\mathbb{E}\{\cdot\}$ denotes statistical expectation. $\mathbf{0}_{M\times N}$ denotes the all-zero matrix with size $M\times N$. ${\rm Tr}\{\cdot\}$ denotes the matrix trace. $\overline{\theta}\!=\!1-\theta$. ${\rm diag}\!=\!\{\cdot\}$ denotes a diagonal matrix with the enclosed elements as its diagonal elements. $[\cdot]^+=\max\{\cdot,0\}$ denotes the maximum between the value in the argument and zero.
\vspace{-0.25cm}
\section{System Model and Proposed Scheme}
\vspace{-0.1cm}
\subsection{System Model and Assumptions}
\vspace{-0.05cm}
We assume a wireless network composed of a legitimate transmitter (Alice), her RF-EH legitimate receiver (Bob), and an eavesdropping node (Eve). Both Alice and Eve are assumed to have {\bf reliable} power supplies. All nodes are equipped with a single antenna. The total
channel bandwidth of $W$ Hz is divided into $N$ orthogonal sub-channels by
using OFDM with each sub-channel experiencing slow fading.  The total transmit power
of Alice per time slot (i.e. OFDM block) is constrained by $P$ Watts. We assume equal power allocation among the available sub-channels to reduce the power consumption at both Alice and Bob due to the optimization of $N$ power levels.\footnote{Note that the power allocated to each sub-channel must be known at both the transmitter and its receiver to enable data decoding at the receiver.} It is noteworthy that equal-power allocation among the sub-channels results in almost the same rate as the case of optimizing all power levels using the water-pouring algorithm~\cite{rhee2000increase}.

 We assume a block-fading channel model where the channel
coefficients (taps) remain unchanged during a coherence time duration,
but change randomly from one coherence time duration to another. The thermal noise samples at node $\ell$ are modeled as zero-mean complex circularly-symmetric Gaussian random variables with power $\kappa_\ell$ Watts.
\vspace{-0.2cm}
\subsection{Proposed Scheme}
\vspace{-0.05cm}
Our proposed scheme is implemented jointly by Alice and Bob. Next, we explain the operation at both the transmitter and receiver sides.
\subsubsection{Transmitter Side}
At the transmitter side, Alice converts the frequency-domain signals to time-domain signals using an $N$-point inverse FFT (IFFT), and adds a CP of size $N_{\rm cp}$ samples to the
beginning of every OFDM block. We assume that the
CP length is longer than the delay spread of the channel between Alice and Bob (Eve), denoted by $\nu_{\rm A-B}$ ($\nu_{\rm A-E}$), to eliminate inter-block interference.\footnote{This is a best-case assumption for Eve; otherwise, her rate will be degraded due to inter-block and intra-block interference.} To secure the legitimate transmissions, we assume that Alice transmits AN signals along with her data signals. The AN signal is designed in a way such that its effect will be removed at Bob prior to information decoding as discussed in Section \ref{sec3}. Alice splits her power between data and AN signals. Assuming that a fraction $\theta$ of the total power $P$ is assigned for data transmission, the power assigned for AN transmission is $\overline{\theta} P$. The transmitted signal by Alice is
\begin{equation} \small \small
\begin{split} \small
\label{signalAlice}
\mathbf{s}_{\rm A}=\mathbf{A}_{\rm cp} \mathbf{F}^* \mathbf{x}+\mathbf{Q} \mathbf{z}
\end{split}
\end{equation}
where $ \mathbf{A}_{\rm cp} \in \mathbb{R}^{(N+N_{\rm cp}) \times N}$ is the CP insertion matrix, $\mathbf{x} \in \mathbb{C}^{N\times 1}$ is the data vector, $\mathbf{Q} \in \mathbb{C}^{(N+N_{\rm cp}) \times N_{\rm cp} }$ is the temporal-AN precoder matrix, and $\mathbf{z} \in \mathbb{C}^{N_{\rm cp}\times 1}$ is the AN vector.

Since Eve's instantaneous CSI is assumed to be unknown at Alice, Alice distributes her power assigned for AN isotropically in all directions (i.e. equally among the AN symbols in $\mathbf{z}$). Hence, the AN symbol power is $\overline{\theta} P/N_{\rm cp}$.

\subsubsection{Receiver Side}
  Assume that the overall time duration of an OFDM block is $T_{\rm OFDM}=N_{\rm T} T_{\rm s}$, where $N_{\rm T}=N+N_{\rm cp}$ and $T_{\rm s}$ is the sampling time. The duration from $0$ to $T_{\rm cp}=N_{\rm cp} T_{\rm s}$ seconds represents the CP duration. The duration from $N_{\rm cp} T_{\rm s}$ to $T_{\rm OFDM}$, whose width is $T_{\rm OFDM}-T_{\rm cp}=T_{\rm OFDM}-N_{\rm cp} T_{\rm s}$ seconds, represents the data symbols in the time domain with $N$ samples, as depicted in Fig. 1. We use the whole CP duration to power the receiver. In particular, since the CP samples are typically discarded prior to information decoding, we exploit their energy to power Bob. Moreover, assume that we already discarded the CP. The remainder of the OFDM block in the time domain (before processing) will be divided into two portions, where the receiver splits the data duration into two portiones of durations $\beta (T_{\rm OFDM}-T_{\rm cp})$ and $(1-\beta)(T_{\rm OFDM}-T_{\rm cp})$, respectively. For the first portion of duration $\beta (T_{\rm OFDM}-T_{\rm cp})$, the receiver will employ a PS scheme. For the second portion of duration $(1-\beta)(T_{\rm OFDM}-T_{\rm cp})$, there is no power splitting and the whole signal will be passed through the data processing and decoding circuits. Thus, the energy harvested during this phase is zero. After sampling the OFDM block in the time domain, we get a set of $N$ samples (FFT size is $N$). The samples obtained during the first portion will be multiplied by a fraction of $\sqrt{\rho}$ (hence, the power of these samples will be multiplied/reduced by a factor $\rho$) and the remaining samples will be inserted into the signal processing phase with full~power.
  \color{black}

\begin{figure}
\vspace{-1cm}
	\centering
		  \includegraphics[width=1\columnwidth]{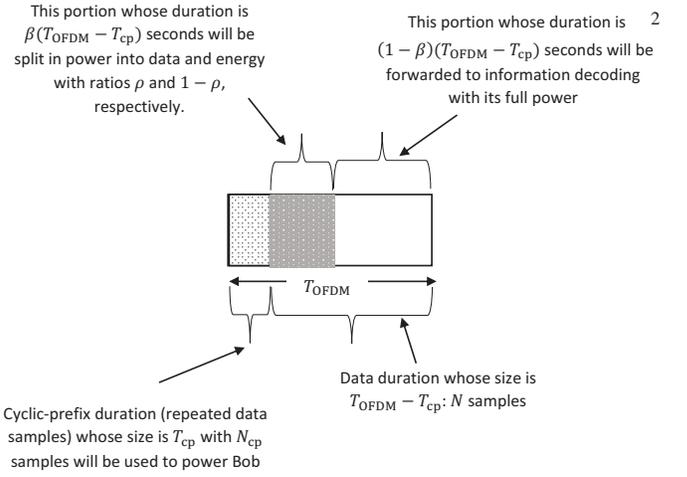}
	\caption{Proposed scheme at the receiver side.}
	\label{figure0}
\vspace{-0.55cm}
\end{figure}

The choice of the CP size affects both the data rate and the energy harvested at the receiver, and thus we will investigate the involved design trade-offs. In practice, the rate is maximized
if $N_{\rm cp}\ge \nu_{\rm A-B}$ is minimized (i.e. $N_{\rm cp}=\nu_{\rm A-B}$). Furthermore, the CP size also impacts the amount
of power that can be scavenged at the receiver. In
particular, the smallest possible $N_{\rm cp}$ might not be optimal in terms of the energy collected at the receiver. Thus, we need to optimize it to further improve the system's energy transfer and secrecy rates.

Since there is no performance gain from using ratios of the time duration $T_{\rm OFDM}$ that are fractions of the sampling time $T_{\rm s}$, $\beta T_{\rm OFDM}$ is assumed to be an integer multiple of $T_{\rm s}$ (symbol duration). Hence, we define a new variable $\gamma \in \{0,1,2,\dots,N\}$ as an integer optimization variable. The value of $\beta$ is obtained from the relation $\beta (T_{\rm OFDM}-T_{\rm cp})= \gamma T_{\rm s}$.
\vspace{-0.25cm}
\section{Secrecy and Energy Transfer Rates} \label{sec3}
\vspace{-0.1cm}
Let $\mathcal{A}$ with cardinality $|\mathcal{A}|=\gamma$ denote the set of time-domain samples taken from $N_{\rm cp}T_{\rm s}$ to $(N_{\rm cp}+\gamma) T_{\rm s}$. Moreover, let $\mathcal{A}^c$ with cardinality $|\mathcal{A}^c|=N-\gamma$ denote the set of time-domain samples taken from $(N_{\rm cp}+\gamma) T_{\rm s}$ to the end of the OFDM block duration. The $i$th received signal sample for $i \in \{1,2,\dots,\gamma\}$, denoted by $y_i \in \mathcal{A}$, is given by
\begin{equation} \small
\begin{split}
y_i= \sqrt{\rho} [ \mathbf{H}_{\rm time} \mathbf{A}_{\rm cp} \mathbf{F}^* \mathbf{x}]_i +\sqrt{\rho}[\mathbf{H}_{\rm time} \mathbf{Q} \mathbf{z}]_i + n_i
\end{split}
\end{equation}
where $\mathbf{H}_{\rm time}$ is the Toeplitz channel matrix between Alice and Bob with the channel-impulse response as its first column, $\mathbf{F}^* \mathbf{x}$ is the IFFT operation on a data vector $\mathbf{x}$, $[\mathbf{F}^* \mathbf{x}]_i$ is the $i$th sample (or element) of the vector $\mathbf{F}^* \mathbf{x}$, and $n_i$ is the $i$th thermal noise sample with variance $\kappa_{\rm B}$.

Sample $j\in \{\gamma+1,\dots,N\}$, denoted by $y_j\in \mathcal{A}^c$, is given~by
\begin{equation} \small
\begin{split}
y_j=    [ \mathbf{H}_{\rm time} \mathbf{A}_{\rm cp} \mathbf{F}^* \mathbf{x}]_j +[\mathbf{H}_{\rm time} \mathbf{Q} \mathbf{z}]_j + n_j
\end{split}
\end{equation}

To obtain the data symbols in the frequency domain, we need first to remove the effect of the power split on the samples in $\mathcal{A}$. Hence, we divide $y_i$ by $\sqrt{\rho}$ and the result is
\begin{equation} \small
\begin{split}
\label{mod}
\tilde{y}_i=y_i/\sqrt{\rho}=  [ \mathbf{H}_{\rm time} \mathbf{A}_{\rm cp} \mathbf{F}^* \mathbf{x}]_i +[\mathbf{H}_{\rm time} \mathbf{Q} \mathbf{z}]_i+ n_i/\sqrt{\rho}
\end{split}
\end{equation}
 The power of the noise sample in \eqref{mod} has been increased by a factor of $1/\rho >1$, where $0\le \rho\le 1$. Then, Bob constructs a vector of observations, $\mathbf{y}=[\tilde{y}_1,\tilde{y}_2,\dots,\tilde{y}_{\gamma},y_{{\gamma}+1},\dots, y_{N}]^\top$, and inserts it into the (digital) FFT block. The result is
\begin{equation} \small
\begin{split}
\label{mod2}
\mathbf{F} \mathbf{y} = \mathbf{H} \mathbf{x} + \mathbf{F} \mathbf{R}_{\rm cp} \mathbf{H}_{\rm time} \mathbf{Q} \mathbf{z} + \mathbf{F}\mathbf{n}
\end{split}
\end{equation}
where the diagonal matrix $\mathbf{H}\!=\!{\rm diag}\left( H_1,H_2,\dots,H_N\right)=\mathbf{F} \mathbf{R}_{\rm cp}\mathbf{H}_{\rm time} \mathbf{A}_{\rm cp} \mathbf{F}^*$ is the channel of the Alice-Bob link in the frequency domain whose $k$th element is the channel coefficient $H_k$ over sub-channel $k$, $\mathbf{R}_{\rm cp} \in \mathbb{R}^{N \times N_{\rm T}}$ is the CP removal matrix, and $\mathbf{n}\in \mathbb{C}^{N\times 1}$ is the noise vector after the CP removal. We note that the variance of each element of $\mathbf{n}$ becomes equal to $\frac{\gamma \frac{\kappa_{\rm B}}{\rho}+(N-\gamma)}{N} \kappa_{\rm B} \ge \kappa_{\rm B}$. This demonstrates that increasing the number of samples that are power split increases the additive noise power (i.e. decreases the received signal-to-noise ratio (SNR) per sub-channel). Hence, as expected, this degrades the data decoding reliability and the achievable data rates as well. However, increasing the fraction of time over which Bob splits his power increases the energy collected by his energy harvester circuit. This represents a trade-off between data and energy transfer~rates.

To cancel the AN at Bob, the AN-precoding matrix at Alice, $\mathbf{Q}$, should satisfy the following condition
\begin{equation} \small
\begin{split}
\label{mod2x}
 \mathbf{R}_{\rm cp} \mathbf{H}_{\rm time} \mathbf{Q} =\mathbf{0}_{N\times N_{\rm cp}}
\end{split}
\end{equation}
Eqn. \eqref{mod2x} has a non-trivial solution since the number of columns of $ \mathbf{R}_{\rm cp} \mathbf{H}_{\rm time} $ is $N_{\rm T}$ and its rank is $N$. Hence, $\mathbf{R}_{\rm cp} \mathbf{H}_{\rm time}$ always has a non-trivial null space.

The achievable rate at Bob is thus given by
\begin{equation} \small
\begin{split}
\label{mod2}
R_{\rm B}=\frac{\log_2 \det\left(\mathbf{I}_{N}+\frac{\theta P}{N_{\rm T}}\mathbf{H} \mathbf{H}^* \left(\mathbf{F}\mathbf{\Sigma_n} \mathbf{F}^*\right)^{-1}\right)}{{N_{\rm T}}}
\end{split}
\end{equation}
where $\mathbf{\Sigma_n}=\kappa_{\rm B}{\rm diag}\{\underbrace{1/\rho,1/\rho,\dots,1/\rho}_{\gamma},1,\dots,1\} \in \mathbb{R}^{N\times N}$.

The processed signal vector at Eve is neither a function of $\rho$ nor $\beta$ ($\gamma$) since these are the legitimate receiver's (Bob's) parameters. Eve's rate is given by
\begin{equation} \small
\begin{split}
\label{mod4}
R_{\rm E}=\frac{\log_2 \det\left(\mathbf{I}_{N}+\frac{\theta P}{N_{\rm T}}\mathbf{G} \mathbf{G}^* \left(\frac{\overline{\theta} P}{N_{\rm cp}}\mathbf{J}\mathbf{J}^*+\kappa_{\rm E} \mathbf{I}_{N}\right)^{-1}\right)}{{N_{\rm T}}}
\end{split}
\end{equation}
where the diagonal matrix $\mathbf{G}={\rm diag}\left( G_1,G_2,\dots,G_N\right)$ is the channel of the Alice-Eve link in frequency domain whose $k$th diagonal element is the channel coefficient $G_k$ over sub-channel $k$, $\mathbf{J}=\mathbf{F} \mathbf{R}_{\rm cp}\mathbf{G}_{\rm time} \mathbf{Q}$, $\mathbf{G}_{\rm time}$ is the Toeplitz channel matrix between Alice and Eve, and $\frac{\overline{\theta} P}{N_{\rm cp}}\mathbf{J} \mathbf{J}^*$ is the AN covariance matrix at Eve.

The system's secrecy rate is given by
\begin{equation} \small
\begin{split}
\label{mod4}
R_{\rm sec}&\!=\! \left[R_{\rm B}\!-\!R_{\rm E}\right]^+ \\& \!=\! \frac{1}{N_{\rm T}}  \left[ {\log_2 \det\left(\mathbf{I}_{N}+\frac{\theta P}{N_{\rm T}}\mathbf{H} \mathbf{H}^* \left(\mathbf{F}\mathbf{\Sigma_n} \mathbf{F}^*\right)^{-1}\right)} \right.\\& \,\,\,\,\,\,\,\,\,\,\,\,\,\,\,\,\,\ \left. - {\log_2 \det\left(\mathbf{I}_{N}\!+\!\frac{\theta P}{N_{\rm T}}\mathbf{G} \mathbf{G}^* \left(\frac{\overline{\theta} P}{N_{\rm cp}}\mathbf{J}\mathbf{J}^*+\kappa_{\rm E} \mathbf{I}_{N}\right)^{-1}\right)} \right]^+
\end{split}
\end{equation}

Bob passes the CP signal to his energy harvester circuit in analog form. The received CP signal is
\begin{equation} \small
\begin{split}
\label{cpret}
\mathbf{q}_{\rm cp} = \mathbf{E}_{\rm cp} \mathbf{H}_{\rm time} \left( \mathbf{A}_{\rm cp} \mathbf{F}^* \mathbf{x}+ \mathbf{Q} \mathbf{z}\right)
\end{split}
\end{equation}
where $\mathbf{E}_{\rm cp} = [\mathbf{I}_{N_{\rm cp}} \ \mathbf{0}_{N_{\rm cp}\times N} ]\in \mathbb{R}^{N_{\rm cp}\times N_{\rm T}}$ is a matrix that extracts $\mathbf{q}_{\rm cp}$, i.e., the portion of $\mathbf{H}_{\rm time} \left( \mathbf{A}_{\rm cp} \mathbf{F}^* \mathbf{x}+ \mathbf{Q} \mathbf{z}\right)$ corresponding to the CP. It should be noted that \eqref{cpret} does not explicitly model the physical operation
performed by the CP retrieval module. It rather provides
a representation of the CP in terms of an equivalent
number of samples $N_{\rm cp}$.

Assuming that the energy conversion efficiency of the energy harvester circuit is $0\le \eta \le 1$, the energy harvested at Bob from the CP is given by
\begin{equation} \small\small
\begin{split}
\label{cpret}
\!\!E_{\rm H,cp}\!=\! \eta {P} {\rm Tr}\left\{\mathbf{E}_{\rm cp} \mathbf{H}_{\rm time}  \left(\frac{\theta \mathbf{A}_{\rm cp}  \mathbf{A}_{\rm cp}^*}{N_{\rm T}} \!+\! \frac{\overline{\theta}\mathbf{Q} \mathbf{Q}^*}{N_{\rm cp}}  \right) \mathbf{H}_{\rm time}^* \mathbf{E}_{\rm cp}^*\right\} T_{\rm cp}
\end{split}
\end{equation}

The energy harvested at Bob from the first $\gamma$ samples after the CP duration is
\begin{equation} \small\small
\begin{split}
\label{modxx}
\!\!\!E_{{\rm H},\gamma}\!=\! \eta \frac{\theta P(1-\rho)}{N_{\rm T}} {\rm Tr}\left\{  \mathbf{K}_\gamma \mathbf{K}^*_\gamma\right\}  \gamma T_{\rm s}
\end{split}
\end{equation}
where $\mathbf{K}_\gamma=\mathbf{E}_{\gamma} \mathbf{R}_{\rm cp}\mathbf{H}_{\rm time} \mathbf{A}_{\rm cp}$, and $\mathbf{E}_{\gamma}=[\mathbf{I}_{\gamma}\  \mathbf{0}_{\gamma\times N-\gamma} ]\in \mathbb{R}^{N_{\rm cp}\times N}$ is a matrix that extracts the $\gamma$ samples corresponding to the time duration where the power is split between energy and information. The total energy harvested at Bob is thus
\begin{equation} \small
\begin{split}
\label{modxx}
E_{\rm H}= E_{\rm H,cp}+ E_{{\rm H},\gamma}
\end{split}
\end{equation}

To reduce the optimization complexity and computational power, and since Eve's CSI is unknown at Alice, we can only optimize the average secrecy rate which requires the Alice-Eve link statistics only. Thus, the optimal solutions of the optimization parameters $\theta$, $\beta$, and $\rho$ are used and kept unchanged throughout the network operation. Our goal is to maximize the average secrecy rate for a given average energy transfer constraint at the receiving node. Our optimization problem is formulated as follows \color{black}
 \begin{equation} \small \small \small
 \begin{split} \small
 \label{OPT}
 \underset{\rho,\gamma,\theta,N_{\rm cp}}{\max:} &\ \mathbb{E}\{R_{\rm sec}\} \\  {\rm s.t.} &  \ \mathbb{E}\{E_{\rm H}\}  \ge \zeta \\ &   \,\ 0 \le \rho,\theta\color{black} \le 1,\  \gamma \in\{0,1,2,\dots,N\} \\& \,\ N_{\rm cp} \in\{\nu_{\rm A-B},\nu_{\rm A-B}+1,\dots,N\} \color{black}
\end{split}
\end{equation}
where $\zeta \ge 0$ is the required/target average energy harvested at the receiver. The optimization problem is nonconvex due to the nonconvexity of the constraints. Alice can solve \eqref{OPT} {\bf OFFLINE} using the Matlab's fmincon function. The solution remains unchanged as long as the system's average parameters (e.g. average channel gains, required average energy at the receiver) remain unchanged. The optimal values of the parameters will be communicated to Bob prior to the actual system's operation. If the optimization problem in \eqref{OPT} is infeasible, the legitimate nodes must agree on a lower value for $\zeta$.

\begin{figure}
   \vspace{-1cm}
  \centering
  \includegraphics[width=0.85\columnwidth]{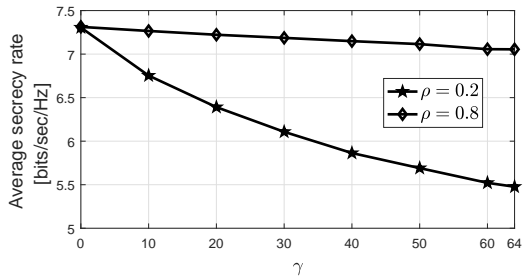}\\
  \caption{Average secrecy rate versus $\gamma$.}
  \label{fig1}
  \vspace{-0.5cm}
  \end{figure}

\vspace{-0.25cm}
\section{Simulation Results}
\vspace{-0.1cm}
We simulate our system using: $2000$ channel realizations, $W=1$ MHz, $N_{\rm cp}=\nu_{\rm A-B}=\nu_{\rm A-E}=16$, $N=64$, $\kappa_{\rm E}=\kappa_{\rm B}=\kappa=1$ Watt, $\eta=0.6$, $\theta=1/2$, and $P/\kappa=40$ dB. Each channel coefficient is assumed to be independent and identically distributed (i.i.d.) zero-mean circularly-symmetric complex Gaussian random variable with unit variance.

\begin{figure}
   \vspace{-1cm}
  \centering
  \includegraphics[width=0.85\columnwidth]{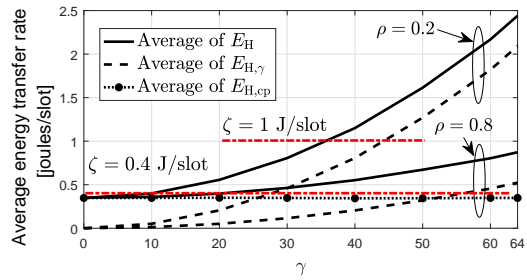}\\
  \caption{Average energy harvested at Bob versus $\gamma$.}
  \label{fig2}
  \vspace{-0.5cm}
  \end{figure}

In Figs. \ref{fig1} and \ref{fig2}, we plot the average secrecy rate and average energy transfer at Bob, respectively, versus $\gamma$ for different values of $\rho$. As shown in the figures, increasing $\gamma$ increases the energy harvested at Bob and decreases the average secrecy rate, as explained earlier. Moreover, increasing $\rho$, which represents the power assigned to data after power splitting at Bob, increases the average secrecy rate and decreases the average energy transfer. Our proposed scheme provides a simple yet efficient approach to control both the average energy transfer rate and the average secrecy rate at Bob. This is because our proposed scheme maximizes the average secrecy rate for a given required average energy transfer rate at Bob.  The case of $\gamma=N=64$ is used as a benchmark to show the gain of our proposed scheme relative to the case when the whole OFDM block is power split into two streams of data and energy with a power splitter. When $\gamma=N$, the average secrecy rate is $5.47$ bits/sec/Hz and $7$ bits/sec/Hz for $\rho=0.2$ and $\rho=0.8$, respectively. Furthermore, the average energy transfer is $0.87$ joules/slot and $2.44$ joules/slot for $\rho=0.8$ and $\rho=0.2$, respectively. If the energy requirement is at least $\zeta=0.4$ joules/slot, and assuming $\rho=0.2$, the minimum value of $\gamma$ to achieve this energy transfer rate is $10$. In our scheme, at $\gamma=10$, the average secrecy rate is $6.75$ bits/sec/Hz. Since the benchmark achieves an average secrecy rate of $5.47$ bits/sec/Hz when $\rho=0.2$, the average secrecy rate gain of our scheme is more than $23\%$. Fig. \ref{fig2} also shows the energy transfer rate due to CP and that due to the second phase of our scheme in which the $\gamma$ samples after the CP removal are also used to energize Bob. The CP energy is independent of $\rho$ and $\gamma$. Moreover, for $\rho=0.2$, the energy harvested from CP samples is higher than the power obtained from the second phase when $\gamma$ is lower than $27$ samples. We also note that if the average energy transfer rate requirement is at least $\zeta=1$ joules/slot instead of $0.4$ joules/slot, $\rho=0.8$ cannot be used since the maximum energy transfer rate when $\rho=0.8$ is $0.87$ joules/slot which happens when $\gamma=N=64$. If $\rho=0.2$, the constraint can be satisfied when $\gamma\ge 36$. In the case of $\gamma=36$, the average secrecy rate is $5.96$ bits/sec/Hz. This demonstrates the importance of optimizing $\rho$ in addition to optimizing $\gamma$.
  \vspace{-0.25cm}
  \section{Conclusions}
  \vspace{-0.1cm}
  We proposed a new secure scheme for OFDM-based SWIPT systems which maximizes the average secrecy rate of a legitimate system under a certain average energy transfer rate at Bob. Our numerical results showed that our proposed scheme achieves an average secrecy rate gain of more than $23\%$ relative to the pure PS scheme.
\vspace{-0.1cm}
\bibliographystyle{IEEEtran}
\vspace{-0.1cm}
\balance
 \bibliography{IEEEabrv,references}
\end{document}